\documentclass[a4paper]{article}

\usepackage{lscape}
\usepackage{amsfonts}
\usepackage{mathrsfs}
\usepackage{xcolor}
\usepackage{cite}
\usepackage{hyperref}

\def\Z{{\mathbb Z}}
\def\bea{\begin{eqnarray}}
\def\eea{\end{eqnarray}}
\def\nn{\nonumber}
\def\half{\frac{1}{2}}

\def\lmatrix{\left(\begin{array}}
\def\rmatrix{\end{array}\right)}
\def\gsim{\mathrel{\rlap{\lower4pt\hbox{\hskip1pt$\sim$}}\raise1pt\hbox{$>$}}}
\def\lsim{\mathrel{\rlap{\lower4pt\hbox{\hskip1pt$\sim$}}\raise1pt\hbox{$<$}}}
\def\bi{\begin{itemize}}
\def\ei{\end{itemize}}
\def\msbar{\overline{\rm MS\kern-0.5pt}\kern0.5pt}
\def\rho{\varrho}
\def\eps{\varepsilon}

\begin{document}

\vspace{2cm}

\begin{center}

    {\Large\bf Dilaton in scalar QFT: a no-go theorem \\ \vspace{0.3cm} in $4-\varepsilon$ and
    $3-\varepsilon$ dimensions}

\vspace{2cm}

Daniel Nogradi and Balint Ozsvath 

\vspace{1cm}

{\em
Eotvos University, Department of Theoretical Physics \\ Pazmany Peter setany 1/a, Budapest 1117, Hungary
}

\end{center}

\vspace{2cm}

\begin{abstract}
Spontaneous scale invariance breaking and the associated Goldstone boson, the dilaton, 
    is investigated in renormalizable, unitary, interacting non-supersymmetric
    scalar field theories in $4-\varepsilon$ dimensions.
    At leading order it is possible to construct models which give rise
    to spontaneous scale invariance breaking classically and indeed a massless dilaton can be identified. Beyond
    leading order, in
    order to have no anomalous scale symmetry breaking in QFT, the models need to be defined at a
    Wilson-Fisher fixed point with exact conformal symmetry. It is shown that
    this requirement on the couplings is incompatible with having the type of flat direction
    which would be necessary for an exactly massless dilaton.
    As a result spontaneous scale symmetry breaking and an exactly massless dilaton
    can not occur in renormalizable, unitary $4-\varepsilon$ dimensional scalar QFT.
    The arguments apply to $\phi^6$ theory in $3-\varepsilon$ dimensions as well.

\end{abstract}

\newpage

\tableofcontents

\section{Introduction}
\label{introduction}

A massless (or approximately massless) dilaton is thought to arise in a number of field theories where
(approximate) scale invariance is spontaneously broken. If this happens an effective field theory should
be able to describe the low energy dynamics of the dilaton along with other potentially light degrees of
freedom. These other light degrees of freedom are often also Goldstone bosons originating from
spontaneous breaking of other symmetries besides scale invariance. These other symmetries 
might be also approximate and in this case the corresponding Goldstones would also be only 
approximately massless.

Conformal invariance, or invariance only by a subgroup, scale transformations, 
may play a role in the Standard Model and its extensions in a number of ways. The only scale symmetry
breaking parameter of the Standard Model is the Higgs mass, at least at leading order. Quantum effects
lead to additional scale symmetry breaking (dimensionful parameters), most notably by anomalous breaking
of scale invariance and by spontaneous breaking of global symmetries. One class of ideas explores the
possibility that scale symmetry itself is spontaneously broken and to what extent the Higgs boson (or
other light particles) may be identified with the corresponding light dilaton 
\cite{Yamawaki:1985zg,Bardeen:1985sm,Bando:1986bg,Clark:1986gx,Dzhikiya:1986kk,Appelquist:2010gy,Grinstein:2011dq,Matsuzaki:2012gd,Bellazzini:2012vz,CPR,Bellazzini:2013fga,Coradeschi:2013gda}.
In particular, the smallness 
of the ratio between the Higgs mass and the Planck scale, which appears to be an enormously fine tuned
quantity, is then explained by scale symmetry and its spontaneous breaking \cite{Bardeen:1995kv}.
The same could be said about another fine tuning problem of the Standard Model, the smallness of the
cosmological constant as well \cite{Wetterich:2019qzx}. For early appearances of spontaneous breaking
of scale invariance and its implications in QFT see \cite{Isham:1970gz, Ellis:1971sa, Fubini:1976jm}.

The Higgs boson may also be viewed as a generic composite particle in several extensions 
of the Standard Model inspired by strong
dynamics \cite{Weinberg:1975gm, Susskind:1978ms} with or without a dilatonic interpretation. 
In these scenarios the spectrum of a 
strongly interacting new sector is thought
to include a composite light scalar and if so, may be identified with the Higgs. Furthermore, the
lightness may be related to the dilatonic nature of the particle although this interpretation is far from
clear \cite{Holdom:1986ub, Holdom:1987yu}. In any case, the strongly interacting nature of the underlying
theory makes the study of detailed properties of the low energy excitations difficult, and served as the
major motivation for a surge in non-perturbative studies recently 
\cite{Fodor:2012ni,LatKMI:2013dnr,LatKMI:2013bhp,LatKMI:2013usl,LatKMI:2014xoh,Rinaldi:2015axa,LatKMI:2016xxi,Appelquist:2017vyy,Appelquist:2017wcg,Fodor:2017nlp,Appelquist:2019lgk,Fodor:2019vmw,Witzel:2019jbe,Fodor:2020niv,Appelquist:2020bqj,Golterman:2020tdq} as well as descriptions in
terms of effective theories \cite{Golterman:2016lsd,LSD:2018inr,Golterman:2018mfm}.

Whether a particular light degree of freedom in a given field theory can be identified as a dilaton is
often non-trivial. Even though the effective theory might be weakly coupled, the underlying theory is
often strongly coupled, complicating the identification of the physical degrees of freedom between the
two. The fact that the light particles in question, one of which would be the hypothetical dilaton, are often 
not exactly massless further complicates the precise identification, especially if there are no
parameters which would control the masses separately. For example, non-abelian gauge theories are often
thought to give rise to a light dilaton if the fermion content is not far from the conformal window, but
the lack of an obvious control parameter for its mass makes this conclusion more conjectural than firmly
established. In this example the underlying theory is strongly coupled so the bridge from it to the
effective theory is beyond analytical tools. Hence it is not possible to simply derive the effective
theory in a top-down approach. As is often the case with effective theories one needs to consider the
most general model with the given degrees of freedom and symmetries and match the unknown coefficients to
observables in the underlying theory. However, if it is not known whether the underlying theory does give
rise to a dilaton or not, one will not know whether a dilaton field should be included in the
effective theory to begin with. Even if a dilaton is included in the effective theory, its interactions
are not sufficiently constrained by scale symmetry. In contrast, the effective theory describing
Goldstones corresponding to chiral symmetry breaking in gauge theory is essentially unique and fixed by
the pattern of symmetry breaking. If a dilaton is to be coupled to these Goldstones, the form of the
coupling is not fixed by scale symmetry and there are different conjectures for the detailed form of the
coupled system.

The main motivation for the present work was to construct a renormalizable, weakly coupled and unitary theory in which
spontaneous scale symmetry breaking unambiguously takes place and is fully in the realm of perturbation
theory. 
In this case it would be possible to follow the dynamics from the underlying 
theory down to the effective theory describing the massless dilaton and 
potentially the other Goldstones, in a fully controlled manner.
In order to avoid special features specific only to supersymmetric theories, non-supersymmetric
models are sought.

Ideally, one would start with a weakly coupled CFT with a vanishing $\beta$-function 
and break scale symmetry spontaneously only, generating massive particles as well as the massless dilaton. 
We know of no renormalizable, non-supersymmetric and unitary such example in 4 dimensions \cite{Bellazzini:2012vz}. 
Hence we will be working in $4-\varepsilon$ dimensions and purely scalar QFT for simplicity,
where the Wilson-Fisher fixed points \cite{Wilson:1971dc} provide the necessary starting point as a weakly coupled CFT.

The main result is that the two essential features of the interaction are incompatible, 
namely that (1) spontaneous scale symmetry breaking is present (2) the model is renormalizable in the UV and
an IR fixed point also exists, can not be met simultaneously.
Hence in renormalizable unitary scalar QFT in $4-\varepsilon$ dimensions 
spontaneous scale symmetry breaking can not take place and consequently a massless dilaton can not exist.

It is well known that supersymmetric models with the desired properties do exist, even in 4 dimensions. 
Most notable are
${\mathscr N} = 2, 4$ SUSY Yang-Mills \cite{Fayet:1978ig}. If the scalar vevs are all zero, scale symmetry
(even the larger full conformal group) is intact and in QFT all particles are massless with a vanishing
$\beta$-function. Giving non-zero vevs to the scalars leads to spontaneous scale symmetry breaking and a
corresponding exactly massless dilaton, along with other massless and massive states. 
Supersymmetry ensures that the flat direction required for the dilaton is not lifted by 
quantum corrections at any order; see e.g. \cite{Schwimmer:2010za}.

Recently, the first non-supersymmetric 4 dimensional example was found \cite{Karananas:2019fox}. In this example the
double scaling limit of $\gamma$-deformed ${\mathscr N}=4$ SUSY Yang-Mills 
\cite{Frolov:2005dj, Fokken:2013aea, Fokken:2014soa, Gurdogan:2015csr} is considered at strong
deformation and weak coupling, in the large-$N$ limit, leading to so-called fishnet CFT's
\cite{Grabner:2017pgm, Karananas:2019fox}.
The resulting theory is purely bosonic and renormalizable, however it is non-unitary. 
Non-unitarity is in fact crucial to show that the flat
direction giving rise to a massless dilaton is not lifted by quantum effects. 

These observations about supersymmetric and non-unitary non-supersymmetric examples were our primary
motivation for the present work.
Our result shows that a well-defined dilaton in renormalizable, non-supersymmetric and unitary perturbative QFT 
is not easy to construct; in purely scalar QFT it is in fact impossible in $4-\varepsilon$ dimensions. 
For a recent review on the fate
of scale symmetry in QFT along with phenomenological applications in cosmology, gravity and particle physics
in general, see \cite{Wetterich:2019qzx}.

The interaction in $4-\eps$ dimensions is of course quartic. The arguments given in this case can be
applied to $\phi^6$ theory in $3-\eps$ dimensions as well. The leading order 2-loop $\beta$-function has
a zero at small $O(\eps)$ coupling, just as with the Wilson-Fisher fixed point in $4-\eps$ dimensions.
The couplings at the fixed points correspond to a potential which does not have a flat direction
necessary for spontaneous scale symmetry breaking and hence a massless dilaton.

In section \ref{dilatonatleadingorder} a number of examples are provided at leading order demonstrating
that a classical dilaton can easily arise, either as the sole massless mode or together with other
Goldstone bosons. Section \ref{radiativecorrections} deals with the 1-loop $\beta$-functions of each
example and is shown that an IR fixed point can not be reached in the available space of couplings. The
general argument is given in section \ref{genericscalarqft} why the presence of an IR fixed point
precludes the scalar potential from having a flat direction necessary for spontaneous scale symmetry breaking
and a massless dilaton in $4-\eps$ dimensions. In section \ref{genericscalarqft3} the same is shown in
$3-\eps$ dimensions with $\phi^6$ theory.
We end with a set of conclusions and future outlook
in section \ref{conclusionandoutlook}.

\section{Dilaton at leading order in 4 dimensions}
\label{dilatonatleadingorder}

We are seeking interacting renormalizable scalar field theories described by scale invariant actions
with the property that this symmetry breaks spontaneously only. Classically, there is no problem with
working in 4 dimensions for illustrating the tree-level structure.
The generic form at tree level is,
\bea
\label{lag}
{\mathscr L} = \frac{1}{2} \partial_\mu \phi_a \partial^\mu \phi_a - {\mathscr V}(\phi)\;,
\eea
where the potential ${\mathscr V}$ contains only dimensionless couplings.
No particular global symmetry is imposed, only scale symmetry, leading to
\bea
\label{v}
{\mathscr V}(\phi) = \frac{\lambda_{abcd}}{4!}  \phi_a \phi_b \phi_c \phi_c\;.
\eea
The real couplings $\lambda_{abcd}$ and real fields $\phi_a$ are understood to be bare quantities. Scale invariance
is clearly present, the action $S=\int d^4x {\mathscr L}$ is invariant under 
the space-time symmetry $x \to e^{-s} x$ once the
scalar field $\phi_a$ is transformed according to its mass dimension, $\phi_a(x)\to e^{s} \phi_a( e^s x)$.
In particular linear, quadratic or cubic terms in the fields are not allowed.

The potential ${\mathscr V}$ is required to be non-negative for stability of the vacuum. Then clearly
$\phi_a=0$ always corresponds to a vacuum state, one which does not break scale invariance and all
particles are massless. We would like
to construct potentials which possess non-trivial minima $\phi_a = v_a \neq 0$. Once such a minimum exists
the corresponding vacuum will break scale invariance and some states will acquire masses proportional to
$v_a$. 

There is no obstruction at leading order, the simplest example is given by a two component model,
\bea
\label{v12}
{\mathscr V} = \frac{\lambda}{4} \phi_1^2 \phi_2^2\;.
\eea
Clearly the potential is non-negative and possesses infinitely many stable minima. The choice
$(\phi_1,\phi_2)=(0,0)$
corresponds to a scale symmetric vacuum, while $(\phi_1,\phi_2) = (v,0)$ and 
$(\phi_1,\phi_2)=(0,v)$, with
arbitrary $v$, correspond to vacua which break scale invariance spontaneously. 

Expanding around the scale invariant vacuum leads to two massless bosons interacting through a quartic
interaction. More interesting is the expansion around a scale symmetry breaking vacuum, for definiteness
let us choose $(\phi_1, \phi_2) = (0,v)$, and the corresponding fluctuating fields will be denoted
by $\eta$ and $\chi$,
\bea
\phi_1 &=& \eta \nn \\
\phi_2 &=& v + \chi\;.
\eea
The potential becomes,
\bea
{\mathscr V} = \frac{\lambda}{4} v^2 \eta^2 + \frac{\lambda}{2} v \eta^2 \chi +
\frac{\lambda}{4} \eta^2 \chi^2\;, 
\eea
which clearly describes a massive particle $\eta$ with $M^2 = \half \lambda v^2$ and a
massless particle $\chi$, the dilaton
\footnote{In this particular example a global $\Z_2 \times \Z_2$ given by flipping the
sign of the two fields is also broken to $\Z_2$ but this is of no significance for our discussion. Same
applies to the models (\ref{phiaphi}) and (\ref{phiphi}).}. 
The two types of particles are interacting through a cubic and quartic interaction.

Goldstone's theorem applies, the direction given by $\chi$ is a flat direction of the potential 
and hence corresponds to a
massless particle. It is well-known that spontaneous breaking of space-time symmetries behave
differently from spontaneous breaking of global symmetries in terms of counting Goldstone bosons, both in
non-Lorentz invariant \cite{Nielsen:1975hm} and in Lorentz invariant theories \cite{Low:2001bw}. In
the case of scale symmetry the naive counting however does apply,
one spontaneously broken symmetry corresponds to one Goldstone boson.

It is possible to generalize the model (\ref{v12}) to describe an interacting $n$-component and a 1-component field,
$\phi_a$ and $\Phi$. The potential
\bea
\label{phiaphi}
{\mathscr V} = \frac{h_{abcd}}{4!} \phi_a \phi_b \phi_c \phi_d + \frac{1}{2} g_{ab} \phi_a \phi_b \Phi^2
\eea
with dimensionless couplings $h_{abcd}$ and $g_{ab}$ fulfilling suitable stability conditions
gives rise to a scale symmetry respecting vacuum $(\phi_a,\Phi) = (0,0)$ as well as scale symmetry
breaking ones $(\phi_a,\Phi) = (0,v)$ with arbitrary $v$. Expanding around one of these
latter vacua we have fluctuating fields $\eta_a$ and $\chi$,
\bea
\phi_a &=& \eta_a \nn \\
\Phi &=& v + \chi\;,
\eea
leading to the potential,
\bea
{\mathscr V} = \frac{1}{2} g_{ab} \eta_a \eta_b v^2 + g_{ab} v \eta_a \eta_b \chi +
\frac{1}{2} g_{ab} \eta_a \eta_b \chi^2 + \frac{h_{abcd}}{4!} \eta_a \eta_b \eta_c \eta_d\;.
\eea
Before symmetry breaking we had $n+1$ massless particles. After scale symmetry breaking we have $n$
massive particles with mass matrix $M^2_{ab} = g_{ab} v^2$ and a massless dilaton described by
$\chi$ and again the two types of particles are interacting through cubic and quartic interactions.
Goldstone's theorem applies again, one spontaneously broken symmetry corresponds to one massless
Goldstone boson.

Something curious is nonetheless going on relative to generic field theories with Goldstone bosons.
Generically, without symmetry breaking particles are massive and Goldstone's theorem provides 
an explanation why some particles {\em become} massless once the symmetry is spontaneously broken. 
In our case, since we start with a scale invariant action, all particles are massless from the start if
scale symmetry is intact. Hence massless particles do not require a special explanation, their vanishing
mass is
simply a consequence of intact scale symmetry. After scale symmetry breaking a mass scale is generated and
 Goldstone's theorem ensures that one particle {\em remains} massless. At the same time all other
particles acquire a mass. The non-trivial content of Goldstone's theorem
in this case seems to be the precise number of particles which become {\em massive}, 
as opposed to becoming {\em massless}, after symmetry breaking.

As a third and final example let us incorporate spontaneous global symmetry breaking along with scale symmetry
breaking, in which case we expect a dilaton as well as other Goldstone bosons. A potential with these
properties is,
\bea
\label{phiphi}
{\mathscr V} = \frac{\lambda}{4!} \left( \phi_a \phi_a - \Phi^2 \right)^2\;.
\eea
The model is clearly scale invariant and also has an $O(n)$ symmetry. 
The trivial vacuum $(\Phi, \phi_a)=(0,0)$ again breaks neither
scale invariance nor the global symmetry, while the non-trivial vacua,
\bea
(\Phi,\phi_1,\ldots,\phi_{n-1},\phi_n)=(v,0,\ldots,0,v)\;,
\eea
with arbitrary $v$ does
break both. In particular $O(n)$ breaks to $O(n-1)$ hence we expect $n$ Goldstones, $n-1$ from the
breaking of the global symmetry and an additional one as the dilaton.
Indeed, if fields $\eta_0, \eta_1, \ldots, \eta_n$ are introduced,
\bea
\Phi &=& v + \eta_0 \nn \\
\phi_A &=& \eta_A, \quad\qquad 1 \leq A \leq n - 1\\
\phi_n &=& v + \eta_n\;, \nn 
\eea
the potential becomes
\bea
{\mathscr V} = \frac{\lambda}{6} v^2 (\eta_n - \eta_0)^2 +
\frac{\lambda}{6}v(\eta_n-\eta_0)(\eta_A^2 + \eta_n^2 - \eta_0^2) + \frac{\lambda}{4!}(\eta_A^2 +
\eta_n^2 - \eta_0^2)^2\;.
\eea
A change of basis to $\xi = (\eta_n-\eta_0) / \sqrt{2}$ and $\chi = (\eta_n + \eta_0) / \sqrt{2}$ then
leads to 
\bea
{\mathscr V} = \frac{\lambda}{3} v^2 \xi^2 +
\frac{\lambda\sqrt{2}}{6}v \xi (\eta_A^2 + 2\xi\chi) + \frac{\lambda}{4!}(\eta_A^2 + 2\xi\chi)^2\;,
\eea
which shows that $\xi$ is massive with $M^2 = \frac{2}{3}\lambda v^2$ and the remaining $n$
particles are massless, $\chi$ is the dilaton and $\eta_A$ are the $n-1$ Goldstones corresponding to the
breaking $O(n) \to O(n-1)$.

Goldstone's theorem applies again, out of the $n+1$ massless particles, exactly
one becomes massive after symmetry breaking, $n$ remain massless.

Model (\ref{phiphi}) may be generalized further to the so-called biconical models. The field $\Phi$
in this case has $m$ components enlarging the original symmetry to $O(n) \times O(m)$. We will not 
discuss this setup further, rather just note that it has recently attracted interest; see
\cite{Chai:2020zgq} for a detailed discussion. 

For completeness we note that the dilaton $\chi$ is often parametrized as 
$\chi = f e^{\sigma / f}$ with a new field $\sigma$ and
dimensionful parameter $f$. Scale transformations $x \to e^{-s} x$ are realized non-linearly on $\sigma$,
\bea
\sigma(x) \to \sigma\left(e^{s} x\right) + f s\;.
\eea

The discussion so far was completely classical and we turn to loop corrections in the next section.

\section{Quantization in $4-\eps$ dimensions}
\label{radiativecorrections}

As shown in the previous section there is no obstruction to unambiguously define an exactly massless dilaton classically
using a suitably chosen potential. One might wonder if a consistent renormalizable QFT can be built
using the corresponding tree-level potentials. In particular we are seeking a non-trivial CFT with vanishing
$\beta$-function. Were such a construction with spontaneous scale symmetry breaking successful, it would provide an
example of a renormalizable interacting QFT describing some massive particles and a
massless dilaton which is not just an effective theory.
The main conclusion from this section will be that this is actually not possible with the examples given in
section \ref{dilatonatleadingorder}. In the next section it will be shown that the same conclusion
applies generally.

Let us work within dimensional regularization and $\msbar$ scheme and all couplings and fields will be
assumed to be renormalized in this section. Since the starting point ought to be a weakly coupled CFT we
work in $4-\eps$ dimensions without anomalous scale symmetry breaking, which would occur in 4 dimensions.
For small $\eps$ the Wilson-Fisher IR fixed points, 
$\lambda_{abcd} = O(\eps)$ are all perturbative. Once $\eps > 0$ the potential in (\ref{lag}) of course
picks up an additional $\mu^\eps$ term, ${\mathscr V} \to \mu^\eps {\mathscr V}$, for dimensional
reasons.

First, let us see the effect of
loop corrections on our simplest model (\ref{v12}). If we are to have the property that a non-trivial
scale symmetry breaking vacuum exists in QFT, $\langle \phi_2 \rangle \neq 0$, 
the form of the potential should either remain the
same as in (\ref{v12}) or only a term of the type $\phi_1^4$ should be generated. In other words the term
$\phi_2^4$ is forbidden. Unfortunately there is no symmetry which would prohibit this term at all loop
order, hence it is expected that it will be generated perturbatively. Indeed, if all terms are included
at tree-level which are generated at 1-loop, so that the theory is renormalizable, we must start 
with the potential,
\bea
{\mathscr V} = \frac{\lambda_1}{4!} \phi_1^4 + \frac{\lambda_2}{4!} \phi_2^4 + \frac{\lambda_{12}}{4} \phi_1^2
\phi_2^2\;.
\eea
The set of renormalized couplings $\lambda_{1}, \lambda_{2}, \lambda_{12}$ do close under the RG flow, at
1-loop we have,
\bea
\label{bbb}
\mu \frac{d\lambda_{1}}{d\mu} &=& - \eps \lambda_1 + \frac{3}{16\pi^2} \left( \lambda_{1}^2 + \lambda_{12}^2 \right) \nn \\
\mu \frac{d\lambda_{2}}{d\mu} &=&  - \eps \lambda_2 + \frac{3}{16\pi^2} \left( \lambda_{2}^2 + \lambda_{12}^2\right)  \\
\mu \frac{d\lambda_{12}}{d\mu} &=&  - \eps \lambda_{12} + \frac{1}{16\pi^2} \lambda_{12} \left( \lambda_{1} + \lambda_{2} + 4
\lambda_{12} \right)\;. \nn
\eea
It is clear what the problem is: the subspace $\lambda_{2} = 0$ is not invariant under the RG flow and
there is no IR fixed point in this plane. The
tree-level potential corresponding to $\lambda_2 = 0$ and $\lambda_1\neq 0, \lambda_{12}\neq 0$ 
has the desired property in terms of giving rise
to spontaneous breaking, but already at 1-loop $\lambda_{2} \neq 0$. Furthermore, the full set of zeros
of the $\beta$-function on the right hand side of equation (\ref{bbb}) with non-zero $\lambda_{12}$ 
can be obtained explicitly. There
are two solutions, either an $O(2)$ invariant model or a model with two decoupled 1-component models.
Neither of these support spontaneous scale symmetry breaking.

This means that the flat direction
which was essential for spontaneous breaking is lifted by quantum effects and a massless dilaton 
is not present.

A similar analysis holds for the model (\ref{phiphi}). In order to include all terms at leading order
which are generated perturbatively, we must consider,
\bea
{\mathscr V} = \frac{\lambda_1}{4!} (\phi_a\phi_a)^2 + \frac{\lambda_2}{4!} \Phi^4 +
\frac{\lambda_{12}}{4} \phi_a \phi_a \Phi^2\;.
\eea
The $\beta$-functions for the three couplings,
\bea
\mu \frac{d\lambda_{1}}{d\mu} &=& -\eps\lambda_1 + \frac{3}{16\pi^2} \left( \frac{n+8}{9} \lambda_1^2 + \lambda_{12}^2  \right) \nn \\
\mu \frac{d\lambda_{2}}{d\mu} &=&  -\eps\lambda_2 +\frac{3}{16\pi^2} \left( \lambda_2^2 + n \lambda_{12}^2 \right)  \\
\mu \frac{d\lambda_{12}}{d\mu} &=&  -\eps\lambda_{12} +\frac{1}{16\pi^2} \lambda_{12} \left( \frac{n+2}{3} \lambda_1 +
\lambda_2 + 4 \lambda_{12} \right)\;. \nn
\eea
again lead to the conclusion that the $\lambda_2 = 0$ subspace is not invariant under the RG flow. And
there are again only two solutions for the zeros of the $\beta$-functions with non-zero $\lambda_{12}$,
namely the $O(n+1)$ invariant model and a decoupled $O(n)$ and 1-component model, neither of which leads
to spontaneous scale symmetry breaking.

It is easy to see that the same reasoning applies to model (\ref{phiaphi}) too.

\section{Generic scalar CFT in $4-\eps$ dimensions}
\label{genericscalarqft}

The conclusions about the three examples also hold generally and is our main result. Either the scalar
potential allows for an IR fixed point if all vevs are zero or the scalar potential has a flat direction
allowing for a massless dilaton if a vev is non-zero, but not both.

Let us start with a generic tree-level potential (\ref{v}) for an
$N$-component real scalar field with real couplings in $4-\eps$ dimensions
and assume the two ingredients necessary for our desired construction,
an IR fixed point for the couplings and a flat direction. The 1-loop $\beta$-function for the
couplings $\lambda_{abcd}$ is,
\bea
\beta_{abcd} = - \eps \lambda_{abcd} + \frac{1}{16\pi^2} 
\left( \lambda_{abef} \lambda_{efcd} + \lambda_{acef} \lambda_{efbd} + \lambda_{adef} \lambda_{efbc}
\right)\;.
\eea
IR fixed points are given by its zeros, namely,
\bea
\label{wf}
 \eps \lambda_{abcd} = \frac{1}{16\pi^2} 
\left( \lambda_{abef} \lambda_{efcd} + \lambda_{acef} \lambda_{efbd} + \lambda_{adef} \lambda_{efbc}
\right)\;,
\eea
and even though these are seemingly simple quadratic equations, a general classification of its solutions
for arbitrary $N$ is still lacking \cite{Osborn:2017ucf,Rychkov:2018vya}. 
In any case the solutions are $\lambda_{abcd} = O(\eps)$.
Let us assume the flat direction, necessary for spontaneous scale symmetry breaking, is given by some
non-zero vev, $\phi_a = v_a$,
\bea
\lambda_{abcd} v_a v_b v_c v_d = 0\;.
\eea
Using (\ref{wf}) this means,
\bea
\lambda_{abef} \lambda_{efcd} v_a v_b v_c v_d = 0\;,
\eea
which implies $\lambda_{abef} v_a v_b = 0$ for real fields and real couplings. Again using (\ref{wf}) we obtain,
\bea
\label{final}
\lambda_{abcd} v_a = 0\;.
\eea

If the field $\phi_a$ is decomposed into parallel and orthogonal components
to $v_a$ then (\ref{final}) means that ${\mathscr V}(\phi)$ can not depend on the parallel
components at all. Which means that the $N$-component model decouples into a free massless 1-component model 
and $N-1$-components with a quartic potential. Continuing the argument
down to $N=1$ we conclude that the only possibility is ${\mathscr V}(\phi)=0$ i.e. no interaction and no
non-trivial IR fixed point, only $N$ independent free massless scalars.

This is the main result of our paper: spontaneous scale symmetry breaking and hence a massless 
dilaton can not arise in $4-\eps$ dimensional scalar CFT. 

\section{Generic scalar CFT in $3-\eps$ dimensions}
\label{genericscalarqft3}

It is possible to generalize the previous section to $3-\eps$ dimensions and $\phi^6$ theory. The
potential in this case is, 
\bea
\label{v6}
{\mathscr V}(\phi) = \frac{\lambda_{abcdef}}{6!}  \phi_a \phi_b \phi_c \phi_d \phi_e \phi_f \;,
\eea
with a totally symmetric coupling $\lambda_{abcdef}$.
The leading order 2-loop $\beta$-function for the couplings is well-known,
\bea
\label{beta6}
\beta_{abcdef} = - \eps \lambda_{abcdef} + \frac{1}{96\pi^2} \left( \lambda_{abcghi} \lambda_{defghi} +
(9\;\; permutations) \; \right)\;,
\eea
where the 9 permutations restore total symmetry in the $abcdef$ indices. 
Wilson-Fisher type fixed points
exist and are solutions of
\bea
\label{wf6}
\eps \lambda_{abcdef} = \frac{1}{96\pi^2} \left( \lambda_{abcghi} \lambda_{defghi} + (9\;\; permutations) \; \right)\;,
\eea
hence $\lambda_{abcdef} = O(\eps)$. See \cite{Osborn:2017ucf} for details on explicit solutions and
higher loop corrections.

Now let us assume the potential (\ref{v6}) has a flat direction and for notational simplicity let this be
$v_a = (0,\ldots,0,v)$. Then ${\mathscr V}(v) = 0$ and using (\ref{wf6}),
\bea
\label{zero1}
0 = \eps \lambda_{NNNNNN} = \frac{10}{96\pi^2} \lambda_{NNNghi} \lambda_{NNNghi}\;,
\eea
which means $\lambda_{NNNghi} = 0$. Let us denote by $\alpha,\beta,\gamma,\ldots = 1,\ldots,N-1$ the directions
orthogonal to the flat direction. In particular we have $\lambda_{\alpha\beta\gamma NNN} = 0$ and also 
$\lambda_{\alpha\beta NNNN} = 0$. Again using (\ref{wf6}) we get,
\bea
\label{zero2}
0 = \eps \lambda_{\alpha\beta NNNN} = \frac{1}{16\pi^2} \lambda_{\alpha NNghi} \lambda_{\beta NNghi}\;, 
\eea
which leads to $\lambda_{\alpha\beta\gamma\delta NN} = 0$ and also in the same fashion as above,
\bea
\label{zero3}
0 &=& \eps \lambda_{\alpha\beta\gamma\delta NN} = \\
&=& \frac{1}{48\pi^2} \left( 
\lambda_{\alpha\beta \zeta\eta\kappa N} \lambda_{\gamma\delta \zeta\eta\kappa N} + 
\lambda_{\alpha\gamma \zeta\eta\kappa N} \lambda_{\beta\delta \zeta\eta\kappa N} + 
\lambda_{\alpha\delta \zeta\eta\kappa N} \lambda_{\gamma\beta \zeta\eta\kappa N} \right)\;. \nn
\eea
This then implies $\lambda_{\alpha\beta\gamma\delta\zeta N} = 0$ 
which is to say that the potential does not depend on the
$N^{th}$ component at all and describes $N-1$ interacting scalar fields and one decoupled free massless
field. Just as in the $4-\eps$ dimensional case the argument can be repeated down to $N=1$, hence a flat
direction is not compatible with a non-trivial fixed point.

\section{Conclusion and outlook}
\label{conclusionandoutlook}

The dynamical appearance of a dilaton and its detailed properties are a non-trivial problem in
QFT. An appealing playground would be a concrete top-down QFT construction which is perturbative, 
non-supersymmetric, renormalizable and unitary, beyond of course the prerequisite spontaneous breaking of scale
invariance itself. In order to have a massless dilaton, scale symmetry should not be broken anomalously,
i.e. the starting point should be a CFT. All 4 ingredients are important: the perturbative nature
of the construction would ensure that all properties can reliably be calculated, the non-supersymmtric
requirement would guarantee that the construction is generic enough, renormaliability would ensure that
the low energy effective theory describing the dilaton and potentially other Goldstones is UV
complete, and unitarity would make sure that the construction has a well-defined Hamiltonian
version.

It might seem at first that fullfilling all 4 requirements would not be difficult, but actually there is
no known example, even if scale symmetry is allowed to be broken anomalously, in 4 dimensions. Anomalous
scale symmetry breaking, as is the case in 4 dimensional scalar QFT, complicates the identification of 
a potentially light dilaton. A rigorous starting point is an exact CFT, in which case an exactly massless
dilaton would emerge provided spontaneous scale symmetry breaking does take place. Hence in this paper $4-\eps$ 
dimensional scalar QFT was considered with couplings which are at an IR fixed point point of the 1-loop 
$\beta$-function. In principle having more than one scalar field components could allow for a flat
direction for the potential leading to spontaneous breaking only, but it turns out there is no potential
with the two properties simultaneously, namely an IR fixed point for the couplings and a flat direction
as well.

Interestingly, in 4 dimensions, the difficulty lies in fullfilling all 4 requirements simultaneously. 
If models are
allowed to be supersymmetric, examples do exist, most notably ${\mathscr N} = 2$ or ${\mathscr N} = 4$
SUSY Yang-Mills theories \cite{Fayet:1978ig}. If renormalizability is dropped, one may choose specific
models from a large class of effective theories; see \cite{Mooij:2018hew,Gretsch:2013ooa,Shaposhnikov:2008xi}
for recent developments. If unitarity is dropped, again examples are known in the
form of fishnet CFT's \cite{Grabner:2017pgm, Karananas:2019fox}. 
If the interaction is allowed to be strong, beyond the realm of perturbation
theory, much less is known rigorously but one may argue that gauge theories with sufficiently large
fermion content could serve as examples, at least for an approximately massless dilaton. Although in the
gauge theory setup the possibility of an exactly massless dilaton was recently investigated 
\cite{DelDebbio:2021xwu}.

The gauge theory situation was part of the motivation for our study. Even if it is accepted that an
approximately massless dilaton
is present in the spectrum, it is not at all clear what the effective theory is describing the coupled
system of Goldstone bosons from chiral symmetry breaking and the dilaton. Consequently it is not at all
clear what the various couplings are and what the detailed properties of the dilaton itself, e.g. its potential,
is. A generic top-down model with calculable properties would probably have shed some light on 
some of these details.

It is still possible that non-supersymmetric, renormalizable, unitary and perturbative theories do exist
with spontaneous scale symmetry breaking in $4$ or $4-\eps$ dimensions. One certainly needs to look beyond purely
scalar QFT's and we hope to address larger classes of models in the future. It may also be the case that
such theories do not exist, in which case a general proof would be desirable.

The arguments about the lack of spontaneous scale symmetry breaking in scalar QFT in $4-\eps$ dimensions 
can easily be applied to $3-\eps$ dimensions and $\phi^6$ theory as well. The results are the same, either the
potential has a flat direction or its couplings are at a small $O(\eps)$ IR fixed point, but not both.

In principle one could investigate the same question in $6-\eps$ dimensions and $\phi^3$ theory, but in
this case the potential is not bounded from below and a stable vacuum can not be defined, hence the
physical meaning of a flat direction, even if exists, is questionable. Note that also from a purely CFT point
of view there appears to be a qualitative difference between the $3-\eps$ and $4-\eps$ dimensional cases on
the one hand and the $6-\eps$ dimensional case on the other \cite{Basu:2015gpa}.

\section*{Acknowledgments}

DN would like to acknowledge very illuminating conversations with 
Csaba Csaky, Gergely Fejos, Sandor Katz, Agostino Patella, Slava Rychkov and Zsolt Szep.

\end{document}